\documentclass[aps,prb,amsmath,amssymb,twocolumn,superscriptaddress]{revtex4}

\usepackage{graphicx}% Include figure files
\usepackage{dcolumn}% Align table columns on decimal point
\usepackage{bm}% bold math

\begin{document}

%{\bf PREPRINT:} 

\title{On the applicability of non-resonant artificial diamagnetics}

\author{L. Jelinek}
\email{jelinel1@fel.cvut.cz} 
\affiliation{Department of Electromagnetic Field, Czech Technical University in Prague, 16627 Prague, Czech Republic}

\author{M. Lapine}
%\email{mlapine@physics.usyd.edu.au} 
\affiliation{CUDOS, School of Physics, The University of Sydney, NSW 2006, Australia}

\author{R. C. McPhedran}
%\email{ross@physics.usyd.edu.au} 
\affiliation{CUDOS, School of Physics, The University of Sydney, NSW 2006, Australia}

%\date{\today}

%\pacs{42.70.Qs, 41.20.Jb, 42.25.Bs, 41.20.-q}


\begin{abstract}
Artificial diamagnetics are prominent for achieving extraordinarily strong diamagnetism in a wide frequency range. However, as far as the magnetic fields outside the artificial medium are concerned, bulk conductors show a very similar pattern. The question arises whether the complicated internal structure of artificial diamagnetics can, to this end, be replaced by a simpler object. We analyse the figure of merit for the application of diamagnetics in magnetic levitation, and show that for an electrically small body, any internal structuring makes the figure of merit worse than that of a simple conducting object. 
\end{abstract}

\maketitle

\section{Introduction}

The idea of creating artificial diamagnetics is more than a century old. It goes back to the work of Weber \cite{Weber-1852}, who realised that a closed metallic loop exhibits diamagnetic properties. However, the magnitude of the magnetic polarisability of a simple closed loop is very limited. A way to overcome this difficulty is described, for example, by Schelkunoff and Friis \cite{Shelkunoff}:  a lumped capacitor inserted into the loop, making a resonant system,  where the diamagnetic properties are greatly enhanced in the vicinity of the resonance frequency. This preliminary design was improved by Hardy and Whitehead \cite{Hardy}, using a distributed capacitance, and was later made popular by Pendry \cite{Pendry-1999} in planar technology. From that point, artificial diamagnetics made of resonant rings became one of the core topics in metamaterials, and various designs have been systematically studied \cite{Marques}. 

Although the resonant behaviour allows for very low polarisability, the price to pay is the narrow frequency band in which the effect exists. Furthermore, implementation at either very low or very high frequencies is limited by the available capacitances. Non-resonant diamagnetics thus still remain attractive. Recently, this problem has been revisited by Lapine \textit{et al.} \cite{Lapine-2013} and also by Belov \textit{et al.} \cite{Belov-2013}. In the first case \cite{Lapine-2013}, it was shown that a dense hexagonal lattice of closed loops with realistic parameters can deliver effective bulk permeability as low as 0.05. However, this design is necessarily anisotropic, which may pose an obstacle for certain applications. This issue has been avoided \cite{Belov-2013} by using a system of closely packed metallic cubes, which is a design similar to that of Wood \cite{Wood-2007}, yet not involving a superconductor. The system of packed cubes offered bulk permeability values of 0.15 in experimental samples \cite{Belov-2013}.

% In this paper we will show, that even though appealing, the above mentioned non-resonant diamagnetics are constrained by fundamental limits that considerably reduce their applicability.

However, while the designs mentioned above are useful for obtaining a diamagnetic response in the bulk, it turns out that, as far as their influence on the surrounding magnetic fields is concerned, they do not offer an ideal solution. As we show in this paper, a much simpler object (an unstructured good conductor) actually always provides a stronger diamagnetic response.

\section{Qualitative consideration}

There are two kinds of diamagnetism. Classical  atomic diamagnetism originates from the response of bound electrons. This is a weak phenomenon with relative permeability very close to unity (${\mu_{\mathrm{r}}} \gtrsim 0.9996$ for common substances). However, the permeability is almost independent of frequency from DC up to the THz range. The other kind of diamagnetism is connected with a bulk response of conducting bodies excited by time varying magnetic fields, which induce circulating currents that can be written as circulation of magnetisation \cite{Jackson01}, being thus equivalent to it. In this case, the current induction is guided by Faraday's law and is thus dependent on frequency. Such diamagnetism   vanishes at DC, with the exception of a perfect electric conductor or a superconductor. To be more specific, imagine a conducting electrically small body excited by a time harmonic magnetic field ${\bf{B}}$ of angular frequency $\omega$ (time convention ${\mathrm{e}^{{\mathrm{j}}\omega t}}$). The tensor of dipolar magnetic polarisability $\bar \alpha $ of the body is usually defined by ${\bf{m}} = \bar \alpha  \cdot {\bf{B}}$, where ${\bf{m}}$ is the induced magnetic dipole moment. In the case of a good conductor, the components of $\bar \alpha $ can be written as \cite{Landau-8,Jackson01}
%%%%%%%%%%%%%%%%%%%%%%%%%%%%%%%%% Equation 1 %%%%%%%%%%%%%%%%%%%%%%%%%
\begin{equation}
\label{Eq1}
{\alpha_{ij}} \approx \frac{{ - {\mathrm{j}}\omega {C_{ij}}}}{{{R_{ij}} + {\mathrm{j}}\omega {L_{ij}}}},
\end{equation}
%%%%%%%%%%%%%%%%%%%%%%%%%%%%%%%%%%%%%%%%%%%%%%%%%%%%%%%%%%%%%%%%%%%%%
where ${C_{ij}} > 0$ are constants dependent on the shape of the body and where ${R_{ij}} > 0,{L_{ij}} > 0$ are the resistances and self-inductances along different current loops. From \eqref{Eq1}, it is evident that in realistic conductors (${R_{ij}} \ne 0$) the polarisability vanishes for $\omega  \to 0$. On the other hand, at frequencies where $\omega {L_{ij}}/{R_{ij}} \gg 1$, such a body is diamagnetic with a weak frequency dependence.

Based on the above discussion, we claim that:
\begin{emph}
	 {For a passive body of a particular shape and volume, away from the resonance, the lowest magnetic polarisability (the strongest diamagnetism) is achieved when the body is   filled by a good conductor. Any internal structuring of the   body leads to higher polarisability (weaker diamagnetism).}
\end{emph}
This claim is easily understood realising that any structuring leads to current confinement. In the region of the confined current the fields are enhanced and with them the magnetic energy and losses (${L_{ij}}$, ${R_{ij}}$ grows).

\section{A canonical example}

In this section, the statement made above will be presented for a canonical  example of a sphere exposed to a homogeneous magnetic field.

\subsection{A sphere of a homogeneous medium}

Consider a sphere of radius $a$ filled with an isotropic medium with material parameters 
${\varepsilon_1},\;{\mu_1},\;{\sigma_1}$, immersed in a background isotropic medium with parameters 
${\varepsilon_2},\;{\mu_2},\;{\sigma_2}$. 
In order to find a suitable magnetic excitation, we assume the lowest order solution of  the vector wave equation, transversal to the $z$-direction \cite{Stratton}, which can be written as (using spherical coordinates)
%%%%%%%%%%%%%%%%%%%%%%%%%%%%%%%%% Equation 2 %%%%%%%%%%%%%%%%%%%%%%%%%
\begin{equation}
\label{Eq2}
\begin{split}
{{\bf{E}}^{{\mathrm{ext}}}} &=  - {\mathrm{j}}{H_0}\frac{{3{Z_2}}}{2}{{\mathrm{j}}_1}\left( {{k_2}r} \right)\sin \theta {\: {{\boldsymbol \varphi }}_0}\\
{{\bf{H}}^{{\mathrm{ext}}}} &= {H_0}\left[ {{{\bf r}}_0}\frac{3}{{{k_2}r}}{{\mathrm{j}}_1}\left( {{k_2}r} \right)\cos \theta  \right. \\
&+ \left. { {{\boldsymbol \theta }}_0}\frac{3}{{2{k_2}r}} \bigl[ {{{\mathrm{j}}_1}\left( {{k_2}r} \right) - {k_2}r{{\mathrm{j}}_0}\left( {{k_2}r} \right)} \bigr]\sin \theta  \right],
\end{split}
\end{equation}
%%%%%%%%%%%%%%%%%%%%%%%%%%%%%%%%%%%%%%%%%%%%%%%%%%%%%%%%%%%%%%%%%%%%%
where ${Z_2} = \omega {\mu_2}/{k_2}$ is the wave impedance of the background medium, ${k_2}$ is the corresponding wavenumber, ${H_0}$ is the magnetic field at the origin and ${{\mathrm{j}}_n}\left( x \right)$ is the spherical Bessel function of order $n$ \cite{Arfken}. 
It is straightforward to show that for $k_2 r \ll 1$ the exciting field \eqref{Eq2} can be rewritten as (using cylindrical coordinates)
%%%%%%%%%%%%%%%%%%%%%%%%%%%%%%%%% Equation 3 %%%%%%%%%%%%%%%%%%%%%%%%%
\begin{equation}
\label{Eq3}
\begin{split}
{{\bf{E}}^{{\mathrm{ext}}}} &\approx  - {\mathrm{j}}{H_0}\frac{{{Z_2}{k_2}\rho }}{2}\: {{\boldsymbol{\varphi }}_0}\\
{{\bf{H}}^{{\mathrm{ext}}}} &\approx {H_0}{{\bf{z}}_0},
\end{split}
\end{equation}
%%%%%%%%%%%%%%%%%%%%%%%%%%%%%%%%%%%%%%%%%%%%%%%%%%%%%%%%%%%%%%%%%%%%%
which represents a homogeneous $z$-directed magnetic excitation. 
Since the excitation itself is a solution of the vector wave equation in spherical coordinates, then the functional form of the field \eqref{Eq2} will be unperturbed by the presence of the sphere. 
Thus, the field in the presence of the sphere can be assumed in the form 
%%%%%%%%%%%%%%%%%%%%%%%%%%%%%%%%% Equation 4 %%%%%%%%%%%%%%%%%%%%%%%%%
\begin{gather}
\label{Eq4}
%\begin{split}
\left.
\begin{aligned}
{\bf{E}} &=  - {\mathrm{j}}{C_1}\frac{{3{Z_1}}}{2}{{\mathrm{j}}_1}\left( {{k_1}r} \right)\sin \theta \: {{\boldsymbol{\varphi }}_0}\\
{\bf{H}} &= {C_1}\left[ {{\bf{r}}_0}\frac{3}{{{k_1}r}}{{\mathrm{j}}_1}\left( {{k_1}r} \right)\cos \theta  \right. \\
&+ \left. \: {{\boldsymbol{\theta }}_0}\frac{3}{{2{k_1}r}} \bigl[ {{{\mathrm{j}}_1}\left( {{k_1}r} \right) - {k_1}r{{\mathrm{j}}_0}\left( {{k_1}r} \right)} \bigr]\sin \theta  \right]
\end{aligned}
\right|
r < a
\\[5pt]
\left.
\begin{aligned}
{\bf{E}} &= {{\bf{E}}^{{\mathrm{ext}}}} - {\mathrm{j}}{C_2}\frac{{3{Z_2}}}{2}{\mathrm{h}}_1^{\left( 2 \right)}\left( {{k_2}r} \right)\sin \theta \: {{\boldsymbol{\varphi }}_0} \\
{\bf{H}} &= {{\bf{H}}^{{\mathrm{ext}}}} + {C_2} \left[ {{\bf{r}}_0}\frac{3}{{{k_2}r}}{\mathrm{h}}_1^{\left( 2 \right)}\left( {{k_2}r} \right)\cos \theta  \right. \\
&+ \left. \: {{\boldsymbol{\theta }}_0}\frac{3}{{2{k_2}r}}\left[ {{\mathrm{h}}_1^{\left( 2 \right)}\left( {{k_2}r} \right) - {k_2}r{\mathrm{h}}_0^{\left( 2 \right)}\left( {{k_2}r} \right)} \right]\sin \theta  \right]
\end{aligned}
\right|
r > a
%\end{split}
\end{gather}
%%%%%%%%%%%%%%%%%%%%%%%%%%%%%%%%%%%%%%%%%%%%%%%%%%%%%%%%%%%%%%%%%%%%%
where ${Z_1} = \omega {\mu_1}/{k_1}$ is the wave impedance of the sphere and where, in order to find the solution regular for $r = 0,\infty $, we have used Bessel functions for $r < a$ and Hankel functions of the second kind for $r > a$ \cite{Arfken}. 
The unknown constants ${C_1},\;{C_2}$ can be determined from the boundary conditions at $r = a$, i.e. from the continuity of the tangential electric and magnetic field. 

Consider an electrically small sphere (${k_2}a \ll 1$) of ${\varepsilon_1} = {\varepsilon_0},\;{\mu_1} = {\mu_0},\;{\sigma_1} \gg 1$ in a vacuum. 
In this case, it is valid to ask what is the magnetic moment produced by the sphere. 
The magnetic moment results from the induced conduction currents ${J_\varphi} = {\sigma_1}{E_\varphi}$ and can be calculated as
%%%%%%%%%%%%%%%%%%%%%%%%%%%%%%%%% Equation 5 %%%%%%%%%%%%%%%%%%%%%%%%%
\begin{multline}
\label{Eq5}
%\begin{split}
{\bf{m}} = \frac{1}{2}\int\limits_V {\left( {{\bf{r}} \times {\bf{J}}} \right){\mathrm{d}}V}  = 
{{\bf{z}}_0}\frac{{3{\mathrm{j}}{a^2}}}{{{\delta ^2}}}  \cdot {C_1} \cdot V 
\\ 
\times \frac{{3\left( {{k_1}a} \right)\cos \left( {{k_1}a} \right) + \left( { - 3 + {{\left( {{k_1}a} \right)}^2}} \right)\sin \left( {{k_1}a} \right)}}{{{{\left( {{k_1}a} \right)}^5}}},
%\end{split}
\end{multline}
%%%%%%%%%%%%%%%%%%%%%%%%%%%%%%%%%%%%%%%%%%%%%%%%%%%%%%%%%%%%%%%%%%%%%
where $V = 4\pi {a^3}/3$ is the volume of the sphere and \mbox{$\delta  = \sqrt {2/\left( {\omega {\mu_1}{\sigma_1}} \right)}$} is the skin-depth. 
It can be checked that for ${\sigma_1} \to \infty $ the magnetic moment \eqref{Eq5} reads ${\bf{m}} \approx  - {{\bf{z}}_0}3 \cdot V \cdot {H_0}/2$, which is the well known result for a perfectly conducting sphere \cite{Collin}.

Consider another scenario with the sphere filled by a non-conducting diamagnetic medium ${\varepsilon_1} = {\varepsilon_0},\;{\mu_1} < {\mu_0},\;{\sigma_1} = 0$. 
The magnetic moment in this case results from the magnetisation and can be calculated as
%%%%%%%%%%%%%%%%%%%%%%%%%%%%%%%%% Equation 6 %%%%%%%%%%%%%%%%%%%%%%%%%
\begin{multline}
\label{Eq6}
%\begin{split}
{\bf{m}} = \int\limits_V {{\bf{M}}{\mathrm{d}}V}  = \left( {\frac{{{\mu_1}}}{{{\mu_0}}} - 1} \right)\int\limits_V {{\bf{H}}{\mathrm{d}}V} = 
\\
{{\bf{z}}_0}3\left( {\frac{{{\mu_1}}}{{{\mu_0}}} - 1} \right)\frac{{\sin \left( {{k_1}a} \right) - {k_1}a\cos \left( {{k_1}a} \right)}}{{{{\left( {{k_1}a} \right)}^3}}} \cdot {C_1} \cdot V,
%\end{split}
\end{multline}
%%%%%%%%%%%%%%%%%%%%%%%%%%%%%%%%%%%%%%%%%%%%%%%%%%%%%%%%%%%%%%%%%%%%%
which for the relevant case of an electrically small sphere (${k_2}a \ll 1,\;{k_1}a \ll 1$) can be approximated as
%%%%%%%%%%%%%%%%%%%%%%%%%%%%%%%%% Equation 7 %%%%%%%%%%%%%%%%%%%%%%%%%
\begin{equation}
\label{Eq7}
{\bf{m}} \approx {{\bf{z}}_0}\left( {\frac{{{\mu_1}}}{{{\mu_0}}} - 1} \right)\frac{{3{\mu_2}}}{{{\mu_1} + 2{\mu_2}}} \cdot V \cdot {H_0}.
\end{equation}
%%%%%%%%%%%%%%%%%%%%%%%%%%%%%%%%%%%%%%%%%%%%%%%%%%%%%%%%%%%%%%%%%%%%%

\begin{figure}
\centering
\includegraphics[width=0.95\columnwidth]{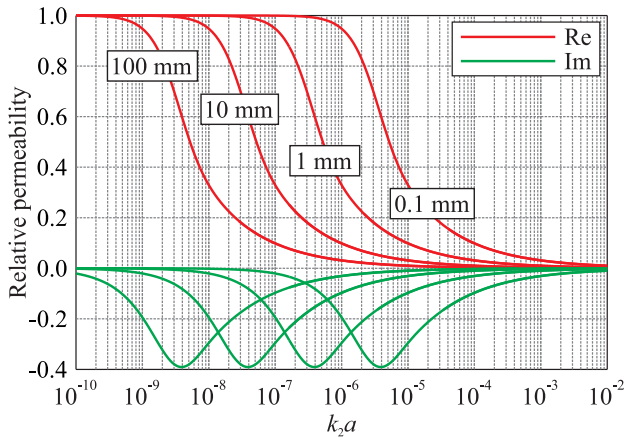}
\caption{\label{fig1} Equivalent permeability for a copper sphere (\mbox{$\sigma  = 5.6 \cdot {10^7}\;{\mathrm{S/m}}$}) for several values of radius $a$. }
\end{figure}

With respect to   artificial diamagnetism it is now interesting to ask what should be the permeability of a diamagnetic sphere such that it would produce the same magnetic moment \eqref{Eq5} as the conducting sphere. This is straightforward to calculate by solving \eqref{Eq7} for $\mu_1$, and the result is depicted in Fig.~\ref{fig1} for a copper sphere. 
%%%%%%%%%%%%%%%%%%%%%%%%%%%%%%%%%%%%%%%%%%%%%%%%%%%%%
%%%%%%%%%%%%%%%%%%%%%%%%%%%%%%%%%%%%%%%%%%%%%%%%%%%%%

Several important observations can be made from Fig.~\ref{fig1}. 
First, the system is not scalable due to the presence of dissipation. 
Second, diamagnetism is only available at frequencies  where the skin-depth  is considerably smaller than the radius of the sphere $a$. 
Third, for frequencies sufficiently high  to overcome dissipation, a good conductor simulates an ideal diamagnetic with ${\mu_{\rm{r}}} \to 0$. 
In fact, this limiting case is evident from \eqref{Eq5} and \eqref{Eq7}, which shows that a PEC sphere corresponds to a sphere filled by ${\mu_1} = 0$. 

It is now instructive to discuss this equivalence also from another perspective. 
Let us calculate the magnetisation current \cite{Jackson01} ${{\bf{J}}_{\bf{M}}} = \nabla  \times {\bf{M}}$ in the ideal diamagnetic case (${\mu_1} \to 0$). Straightforward derivation leads to 
%%%%%%%%%%%%%%%%%%%%%%%%%%%%%%%%% Equation 7a %%%%%%%%%%%%%%%%%%%%%%%%%
\begin{equation}
\begin{split}
{{\bf{J}}_{\bf{M}}} &\approx  - \frac{3}{2}{H_0}\delta \left( {r - a} \right)\sin \theta \: {{\boldsymbol {\varphi }}_0}\\
{I_{\bf{M}}} &\approx - 3{H_0}a,
\end{split}
\label{Eq7a}
\end{equation}
%%%%%%%%%%%%%%%%%%%%%%%%%%%%%%%%%%%%%%%%%%%%%%%%%%%%%%%%%%%%%%%%%%%%%
where ${I_{\bf{M}}}$ denotes the total magnetisation current. The distribution \eqref{Eq7a}  also means that in this case the magnetisation as well as the magnetic field are practically homogeneous inside the sphere. 

On the other hand, the current density   inside the highly conducting  sphere (${\sigma_1} \to \infty$) can also be calculated in a straightforward manner as ${\bf{J}} = {\sigma_1}{\bf{E}}$, which leads to
%%%%%%%%%%%%%%%%%%%%%%%%%%%%%%%%% Equation 7b %%%%%%%%%%%%%%%%%%%%%%%%%
\begin{equation}
\begin{split}
{\bf{J}} &\approx  - \frac{3}{2}{H_0}\left( {{\mathrm{j}}{k_1}a\frac{{{\mathrm{e}^{ - {\mathrm{j}}{k_1}\left( {a - r} \right)}}}}{r}} \right)\sin \theta \: {{\boldsymbol {\varphi }}_0}\\
I &\approx - 3{H_0}a,
\end{split}
\label{Eq7b}
\end{equation}
%%%%%%%%%%%%%%%%%%%%%%%%%%%%%%%%%%%%%%%%%%%%%%%%%%%%%%%%%%%%%%%%%%%%%
where $I$ denotes the total current as in Eq.~\eqref{Eq7a}. In contrast to \eqref{Eq7a}, the current represented by \eqref{Eq7b} leads to a strongly inhomogeneous magnetic field inside the sphere (a strong skin-effect). However, comparing \eqref{Eq7a} and \eqref{Eq7b} it is   important to realise that for ${\sigma_1} \to \infty$ the bracketed term in \eqref{Eq7b} corresponds to the Dirac delta-function, which means that the current distribution in an ideal diamagnetic sphere is   equal to the surface current in the case of a PEC sphere.

\subsection{A sphere of an artificial medium made of conducting loops}

Now we compare the results for the conducting sphere,  presented in Fig.~\ref{fig1}, with those for a sphere made  of an artificial diamagnetic metamaterial --- an anisotropic lattice of closed conducting loops \cite{Lapine-2013}. We consider rings of mean radius $r$, made of a metallic wire (with the same conductivity $\sigma$) of circular cross-section with radius $r_\mathrm{w}$. The rings are arranged in a lattice with periods $b_1 r$ along the axis of the loops, and $b_2 r$ in the plane of the loops. Dimensionless parameters $b_i$ ($i = 1,\,2$) and $w = r_\mathrm{w} / r$, serve for ease of notation. The effective permeability of such a metamaterial can be calculated as \cite{Lapine-2013}
%%%%%%%%%%%%%%%%%%%%%%%%%%%%%%%%% Equation 8 %%%%%%%%%%%%%%%%%%%%%%%%%
\begin{equation}
\mu_{\mathrm{r}} = 1 - \left\lbrack
\dfrac{b_1 b_2^2}{\gamma \pi^2}
\left(
\dfrac{L_\text{e} + \mu_0 r \Sigma}{\mu_0 r} - \dfrac{1}{\zeta} \:
\dfrac{\mathrm{J}_0(\zeta)}{\mathrm{J}_1(\zeta)}
\right)
 + \dfrac{1}{3}
\right\rbrack ^{-1},
\label{Eq8}
\end{equation}
%%%%%%%%%%%%%%%%%%%%%%%%%%%%%%%%%%%%%%%%%%%%%%%%%%%%%%%%%%%%%%%%%%%%%
where {$\zeta = (1 - \mathrm{j}) r w / \delta$, and $\mathrm{J}_i$}  are the $i$-th  order Bessel functions \cite{Arfken}. The total inductance includes the external contribution to the self-inductance of the ring
$L_{\mathrm{e}} = \mu_0 r \left( \ln(8/w) - 2 \right)$ as well as the mutual inductance in the array, reflected by the dimensionless 
$\Sigma = \sum\limits_{n \neq n'} L_{nn'} \bigr/ (\mu_0 r)$, 
which depends on the lattice geometry. 

\begin{figure}
\centering
\includegraphics[width=0.95\columnwidth]{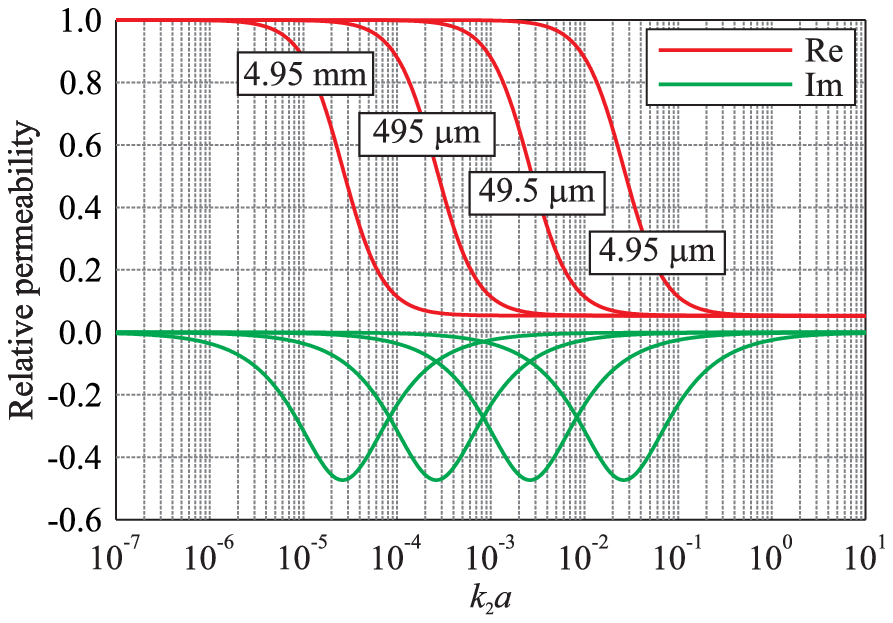}
\caption{\label{fig2} Effective permeability of a metamaterial made of close copper loops (\mbox{$\sigma  = 5.6 \cdot {10^7}\;{\mathrm{S/m}}$}) with different radii $r$ and the same relative geometry and lattice. The radius of the ring is set so that there are 10 unit cells per radius of the corresponding sphere in Fig.~\ref{fig1}.}
\end{figure}

To be consistent with effective medium approximation, in comparing a piece of diamagnetic metamaterial with the metallic sphere, we require $b_i r \ll a$. For a practical example, we choose a lattice constant ten times smaller than the radius of the sphere. At the same time, we know that $b_i$ should be as small as possible to achieve a strong diamagnetism, while $w$ should be as large as possible for the given lattice parameters \cite{Lapine-2013}. Based on the available data (see Fig.~3, Fig.~4 in Ref.~\cite{Lapine-2013}), we select a shifted hexagonal lattice and set $w = 0.01$, $b_1 = 0.02$, $b_2 = 2.02$, as a compromise between sufficiently low permeability and technological constraints (indeed, a further decrease in these parameters does not lower   the $\mu$ values perceptibly). Fig.~\ref{fig2} shows the effective bulk permeability of the ring metamaterial with the above parameters for loops with $r = a / (10 b_2)$ for the same set of $a$ as in Fig.~\ref{fig1}. 
%%%%%%%%%%%%%%%%%%%%%%%%%%%%%%%%%%%%%%%%%%%%%%%%%%%%%
%%%%%%%%%%%%%%%%%%%%%%%%%%%%%%%%%%%%%%%%%%%%%%%%%%%%%

The curves look qualitatively similar to those of a conducting sphere, but the frequency scale in this plot (having the same normalisation) is quite different, indicating that the transition to diamagnetism in this  metamaterial  occurs at a much higher frequency. Also, we note that, in contrast to the equivalent permeability calculated for the conducting sphere, the effective permeability of the lattice of rings does not reach zero.

In this step we could also make a comparison between a bulk conducting sphere and the cubic system reported in Ref.~\onlinecite{Belov-2013}. However a similar result can be expected, since the smallest effective permeability requires the separation between the cubes to vanish, which eventually converges to a bulk metallic body. Introducing a   separation between the cubes will then lead to worse performance.

\section{Magnetic levitation}

To illustrate a practical  consequence of the above arguments, we consider an example relevant for one of the potential applications of artificial diamagnetics: magnetic levitation, which enjoys a fresh attention in the context of metamaterials \cite{Urzhumov2012,Engheta2014}.

When a diamagnetic object of polarisability $\alpha $ is placed into an inhomogeneous magnetic field, a force ${\bf{F}} = \alpha \nabla {\left\| {\bf{B}} \right\|^2}$ tries to expel the object from the field \cite{Stratton}. Magnetic levitation occurs when this magnetic force counterbalances the force of gravity. Imagining that an electrically small levitating object in a vacuum is made of a structured metal (artificial diamagnetics), it is possible to show that for the easiest levitation the quantity ${\mu_0}\left| \alpha  \right| \bigr / \left( {fV} \right)$ should be maximised, with $V$ being the volume to which the polarisability corresponds and $f = {V_{{\rm{metal}}}}/V$ is the volume fraction of metal in that volume. This quantity will hereafter be used as a   figure of merit (FOM) of the diamagnetic properties. Referring to the last section, where the permeability equivalent to the polarisability was derived, it is easy to see that we can further write ${\rm{FOM}} = \left| {\mu_{\rm{r}}^{{\rm{eq}}} - 1} \right|/f$.

%%%%%%%%%%%%%%%%%%%%%%%%%%%%%%%%%%%%%%%%%%%%%%%%%%%%%
\begin{figure}[b]
\centering
\includegraphics[width=0.95\columnwidth]{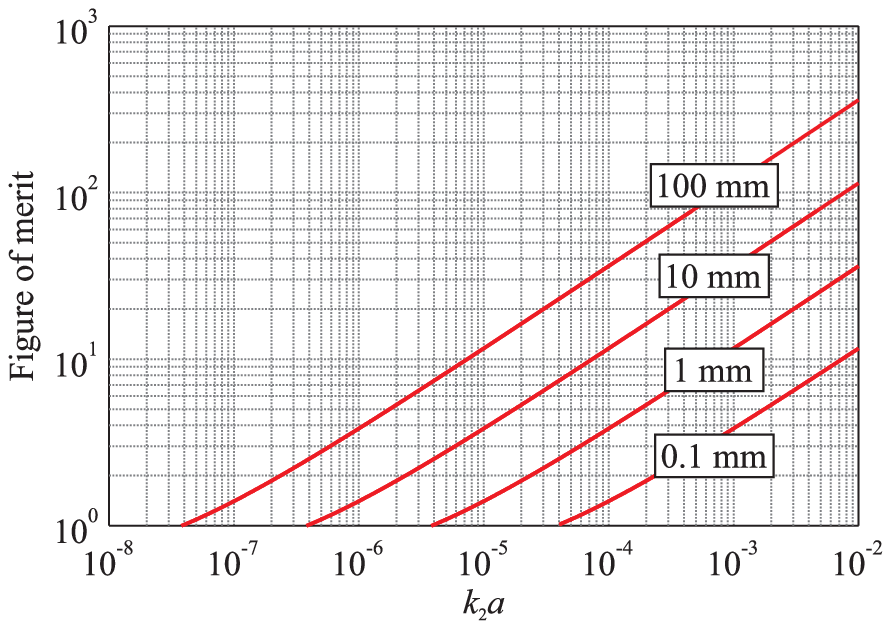}
\caption{\label{fig3} 
Magnetic levitation figure of merit for a copper spherical shell of thickness $3 \delta$ with the same parameters as in 
Fig.~\protect\ref{fig1}. }
\end{figure}
%%%%%%%%%%%%%%%%%%%%%%%%%%%%%%%%%%%%%%%%%%%%%%%%%%%%%

As shown in the previous section, the lattice of rings \cite{Lapine-2013} can offer permeability very close to zero and at the same time it can contain significantly less metal ($f = 2 \pi^2 w^2 r^2 / (b_1 b_2)$  provided that the rings are filled with metal), than a  complete conducting body. It could thus represent a valid competitor to a conducting body with respect to magnetic levitation. 

However, when evaluating the levitation FOM of the conducting sphere it is important to recall that (see the previous section)   diamagnetism only occurs   when the penetration depth is significantly smaller than the radius of the sphere. Therefore   diamagnetic properties result  from a thin surface layer, while the inner part of the conductor can safely be removed. This provides a clear improvement, as the levitation FOM is inversely proportional to the volume fraction $f$. To be conservative, we can take a layer of thickness $3\delta $, which, for the conducting sphere, results in 
%%%%%%%%%%%%%%%%%%%%%%%%%%%%%%%%% Equation 9 %%%%%%%%%%%%%%%%%%%%%%%%%
\begin{equation}
\label{Eq9}
{\rm{FOM}} = \frac{{\left| {\mu_{\rm{r}}^{{\rm{eq}}} - 1} \right|}}{{1 - {{\left( {1 - 3\delta /a} \right)}^3}}}.
\end{equation}
%%%%%%%%%%%%%%%%%%%%%%%%%%%%%%%%%%%%%%%%%%%%%%%%%%%%%%%%%%%%%%%%%%%%%
The result of \eqref{Eq9} is depicted in Fig.~\ref{fig3} for a copper sphere.

%%%%%%%%%%%%%%%%%%%%%%%%%%%%%%%%%%%%%%%%%%%%%%%%%%%%%
\begin{figure}[b]
\centering
\includegraphics[width=0.95\columnwidth]{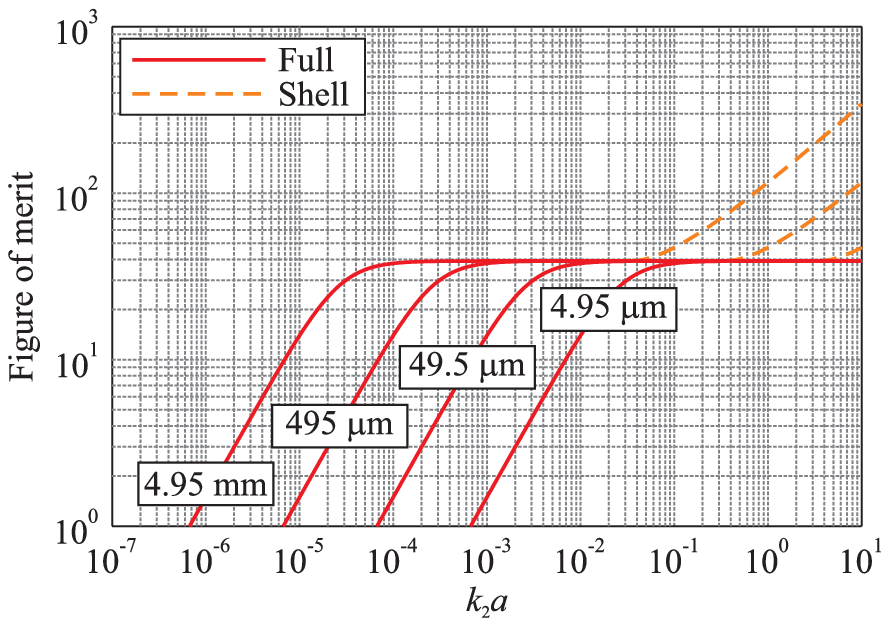}
\caption{\label{fig4} Magnetic levitation figure of merit for a metamaterial made of closed copper loops with the same parameters as in  Fig.~\protect\ref{fig2}. The loops are either complete metal loops (solid lines) or, where possible (when $r_{\mathrm{w}} > 3 \delta$), hollow toroidal shells with wall thickness of $3 \delta$ (dashed lines).  }
\end{figure}
%%%%%%%%%%%%%%%%%%%%%%%%%%%%%%%%%%%%%%%%%%%%%%%%%%%%%

For the  metamaterial made of conducting rings, this approach can also be employed by using hollow toroidal shells with thickness $3 \delta$. However, as apparent from Fig.~\ref{fig4}, for a significant range of the parameters this possibility does not come into play as the rings would actually have to be thinner than the skin-depth.  This fact imposes a saturation of the figure of merit, since the permeability decreases only weakly upon the transition to the diamagnetic regime, while the filling fraction remains the same until the skin-depth becomes smaller than the wire radius (when eventually it starts to increase as shown by the dashed lines).

% This makes a clear difference to the conducting sphere, where the transition to an apparent diamagnetic response occurs at relatively low frequencies, while the skin depth amounts to a small fraction of its radius. 

%%%%%%%%%%%%%%%%%%%%%%%%%%%%%%%%%%%%%%%%%%%%%%%%%%%%%
\begin{figure}
\centering
\includegraphics[width=0.95\columnwidth]{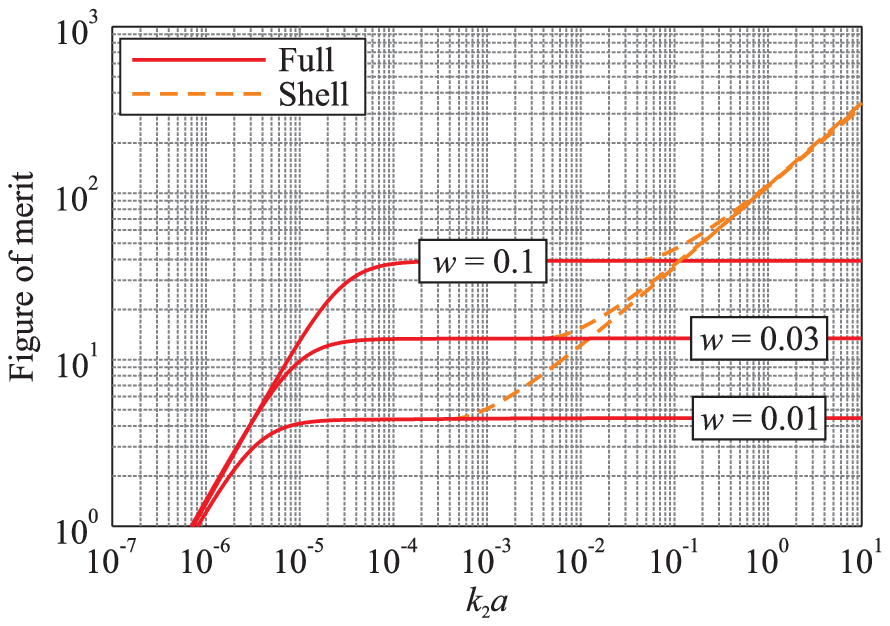}
\caption{\label{fig5} The magnetic levitation figure of merit of a metamaterial made of  either complete metal loops (solid lines) or hollow toroidal shells with wall thickness of $3 \delta$ (dashed lines). The three depicted curves are described by the following sets of parameters 
$\lbrace r = 4.545 \cdot 10^{-3} \mathrm{mm}$, $w = 0.1$, $b_1 = 0.2$, $b_2 = 2.2 \rbrace$,  
$\lbrace r = 4.854 \cdot 10^{-3} \mathrm{mm}$, $w = 0.03$, $b_1 = 0.06$, $b_2 = 2.06 \rbrace$,  
$\lbrace r = 4.95 \cdot 10^{-3} \mathrm{mm}$, $w = 0.01$, $b_1 = 0.02$, $b_2 = 2.02 \rbrace$.  
All the cases obey $r_{\mathrm{w}} > 3 \delta$.}
\end{figure}
%%%%%%%%%%%%%%%%%%%%%%%%%%%%%%%%%%%%%%%%%%%%%%%%%%%%%

This observation leaves the question whether it is possible to increase the figure of merit  by using a thicker hollow wire, so that the relation $3 \delta < r_{\mathrm{w}}$ is valid for a lower frequency. Indeed, the magnitude of $|\mu_{\mathrm{r}} - 1|$ does not change much (from $\sim 0.95$ to $\sim 0.9$) when $w$ is increased ten times and the other parameters are adjusted accordingly, but the filling fraction will benefit from using the skin-effect. %To see the effect, we plot the figure of merit for three sets of parameters as shown in .
 The result (see Fig.~\ref{fig5}), however, is such that although the curves corresponding to the strong skin effect (dashed lines) depart earlier from the saturation range in thicker rings, they all coincide with those for thin rings  at higher frequencies. Thus, the thickness of the rings (within reasonable limits) is irrelevant for the figure of merit.

We also note that for the same reasons the system of separated cubes \cite{Belov-2013} also cannot compete with the conducting body as it has a weaker diamagnetism and a less advantageous filling fraction at the same time.
 
\section{Conclusions}

We have  shown that no passive artificial non-resonant structure can provide a stronger diamagnetic response in the outer region of the body than a commonly available good conductor. A stronger diamagnetic response can be achieved only through resonant or active systems, but at the price of much higher complexity and limited frequency bandwidth.

This conclusion limits practical applications of artificial diamagnetics when the fields outside the body are concerned, in particular, with levitation.   

At the same time, we emphasise that the structure of the fields inside an artificial diamagnetic is quite different from that of a conducting body or shell, and this leaves room for corresponding applications.

We also note that artificial diamagnetics  provide greater freedom in the initial design as well as easy tunability and reconfigurability by changing the lattice. 

Finally, unlike bulk conductors, artificial diamagnetics may possibly be combined with artificial dielectrics or even active inclusions, with an almost independently engineered response, although the mutual interaction between the two sub-systems may be essential and requires a careful analysis.

\begin{acknowledgments}
We thank Pavel A. Belov for useful discussions.
This work has been supported by the Czech Science Foundation (Project No.~13-09086S), by the Czech Technical University in Prague (Project No.~SGS13/198/OHK3/3T/13), and by the Australian Research Council (Project  DP110105484). 
\end{acknowledgments}

\bibliography{References}

\begin{thebibliography}{15}
\expandafter\ifx\csname natexlab\endcsname\relax\def\natexlab#1{#1}\fi
\expandafter\ifx\csname bibnamefont\endcsname\relax
  \def\bibnamefont#1{#1}\fi
\expandafter\ifx\csname bibfnamefont\endcsname\relax
  \def\bibfnamefont#1{#1}\fi
\expandafter\ifx\csname citenamefont\endcsname\relax
  \def\citenamefont#1{#1}\fi
\expandafter\ifx\csname url\endcsname\relax
  \def\url#1{\texttt{#1}}\fi
\expandafter\ifx\csname urlprefix\endcsname\relax\def\urlprefix{URL }\fi
\providecommand{\bibinfo}[2]{#2}
\providecommand{\eprint}[2][]{\url{#2}}

\bibitem[{\citenamefont{Weber}(1852)}]{Weber-1852}
\bibinfo{author}{\bibfnamefont{W.}~\bibnamefont{Weber}}, \bibinfo{journal}{Ann.
  Phys.} \textbf{\bibinfo{volume}{87}}, \bibinfo{pages}{163}
  (\bibinfo{year}{1852}).

\bibitem[{\citenamefont{Schelkunoff and Friis}(1952)}]{Shelkunoff}
\bibinfo{author}{\bibfnamefont{S.~A.} \bibnamefont{Schelkunoff}}
  \bibnamefont{and} \bibinfo{author}{\bibfnamefont{H.~T.} \bibnamefont{Friis}},
  \emph{\bibinfo{title}{Antennas: Theory and Practice}}
  (\bibinfo{publisher}{John Wiley \& Sons, Inc.}, \bibinfo{year}{1952}).

\bibitem[{\citenamefont{Hardy and Whitehead}(1981)}]{Hardy}
\bibinfo{author}{\bibfnamefont{W.~N.} \bibnamefont{Hardy}} \bibnamefont{and}
  \bibinfo{author}{\bibfnamefont{L.~A.} \bibnamefont{Whitehead}},
  \bibinfo{journal}{Rev.\ Sci.\ Instrum.} \textbf{\bibinfo{volume}{52}},
  \bibinfo{pages}{213} (\bibinfo{year}{1981}).

\bibitem[{\citenamefont{Pendry et~al.}(1999)\citenamefont{Pendry, Holden,
  Robbins, and Stewart}}]{Pendry-1999}
\bibinfo{author}{\bibfnamefont{J.~B.} \bibnamefont{Pendry}},
  \bibinfo{author}{\bibfnamefont{A.~J.} \bibnamefont{Holden}},
  \bibinfo{author}{\bibfnamefont{D.~J.} \bibnamefont{Robbins}},
  \bibnamefont{and} \bibinfo{author}{\bibfnamefont{W.~J.}
  \bibnamefont{Stewart}}, \bibinfo{journal}{IEEE T Microw. Theory}
  \textbf{\bibinfo{volume}{47}}, \bibinfo{pages}{2075} (\bibinfo{year}{1999}).

\bibitem[{\citenamefont{Marqu\'es et~al.}(2007)\citenamefont{Marqu\'es,
  Mart\'in, and Sorolla}}]{Marques}
\bibinfo{author}{\bibfnamefont{R.}~\bibnamefont{Marqu\'es}},
  \bibinfo{author}{\bibfnamefont{F.}~\bibnamefont{Mart\'in}}, \bibnamefont{and}
  \bibinfo{author}{\bibfnamefont{M.}~\bibnamefont{Sorolla}},
  \emph{\bibinfo{title}{Metamaterials with Negative Parameters: Theory and
  Microwave Applications}} (\bibinfo{publisher}{John Wiley \& Sons, Inc.},
  \bibinfo{year}{2007}).

\bibitem[{\citenamefont{Lapine et~al.}(2013)\citenamefont{Lapine, Krylova,
  Belov, Poulton, McPhedran, and Kivshar}}]{Lapine-2013}
\bibinfo{author}{\bibfnamefont{M.}~\bibnamefont{Lapine}},
  \bibinfo{author}{\bibfnamefont{A.~K.} \bibnamefont{Krylova}},
  \bibinfo{author}{\bibfnamefont{P.~A.} \bibnamefont{Belov}},
  \bibinfo{author}{\bibfnamefont{C.~G.} \bibnamefont{Poulton}},
  \bibinfo{author}{\bibfnamefont{R.~C.} \bibnamefont{McPhedran}},
  \bibnamefont{and} \bibinfo{author}{\bibfnamefont{Y.~S.}
  \bibnamefont{Kivshar}}, \bibinfo{journal}{Phys.\ Rev.~B}
  \textbf{\bibinfo{volume}{87}}, \bibinfo{pages}{024408}
  (\bibinfo{year}{2013}).

\bibitem[{\citenamefont{Belov et~al.}(2013)\citenamefont{Belov, Slobozhanyuk,
  Filonov, Yagupov, Kapitanova, Simovski, Lapine, and Kivshar}}]{Belov-2013}
\bibinfo{author}{\bibfnamefont{P.~A.} \bibnamefont{Belov}},
  \bibinfo{author}{\bibfnamefont{A.~P.} \bibnamefont{Slobozhanyuk}},
  \bibinfo{author}{\bibfnamefont{D.~S.} \bibnamefont{Filonov}},
  \bibinfo{author}{\bibfnamefont{I.~V.} \bibnamefont{Yagupov}},
  \bibinfo{author}{\bibfnamefont{P.~V.} \bibnamefont{Kapitanova}},
  \bibinfo{author}{\bibfnamefont{C.~R.} \bibnamefont{Simovski}},
  \bibinfo{author}{\bibfnamefont{M.}~\bibnamefont{Lapine}}, \bibnamefont{and}
  \bibinfo{author}{\bibfnamefont{Y.~S.} \bibnamefont{Kivshar}},
  \bibinfo{journal}{Appl.\ Phys.\ Lett.} \textbf{\bibinfo{volume}{103}},
  \bibinfo{pages}{211903} (\bibinfo{year}{2013}).

\bibitem[{\citenamefont{Wood and Pendry}(2007)}]{Wood-2007}
\bibinfo{author}{\bibfnamefont{B.}~\bibnamefont{Wood}} \bibnamefont{and}
  \bibinfo{author}{\bibfnamefont{J.~B.} \bibnamefont{Pendry}},
  \bibinfo{journal}{J. Phys.: Condens. Matter} \textbf{\bibinfo{volume}{19}},
  \bibinfo{pages}{076208} (\bibinfo{year}{2007}).

\bibitem[{\citenamefont{Jackson}(1998)}]{Jackson01}
\bibinfo{author}{\bibfnamefont{J.~D.} \bibnamefont{Jackson}},
  \emph{\bibinfo{title}{Classical Electrodynamics}} (\bibinfo{publisher}{John
  Wiley \& Sons, Inc.}, \bibinfo{year}{1998}), \bibinfo{edition}{3rd} ed.

\bibitem[{\citenamefont{Landau et~al.}(1984)\citenamefont{Landau, Lifshitz, and
  Pitaevskii}}]{Landau-8}
\bibinfo{author}{\bibfnamefont{L.~D.} \bibnamefont{Landau}},
  \bibinfo{author}{\bibfnamefont{E.~M.} \bibnamefont{Lifshitz}},
  \bibnamefont{and} \bibinfo{author}{\bibfnamefont{L.~P.}
  \bibnamefont{Pitaevskii}}, \emph{\bibinfo{title}{Electrodynamics of
  Continuous Media}} (\bibinfo{publisher}{Pergamon Press},
  \bibinfo{year}{1984}), \bibinfo{edition}{2nd} ed.

\bibitem[{\citenamefont{Stratton}(1941)}]{Stratton}
\bibinfo{author}{\bibfnamefont{J.~A.} \bibnamefont{Stratton}},
  \emph{\bibinfo{title}{Electromagnetic Theory}}
  (\bibinfo{publisher}{McGraw-Hill Book Co.}, \bibinfo{year}{1941}).

\bibitem[{\citenamefont{Arfken and Weber}(2005)}]{Arfken}
\bibinfo{author}{\bibfnamefont{G.~B.} \bibnamefont{Arfken}} \bibnamefont{and}
  \bibinfo{author}{\bibfnamefont{H.~J.} \bibnamefont{Weber}},
  \emph{\bibinfo{title}{Mathematical Methods for Physicists}}
  (\bibinfo{publisher}{Elsevier}, \bibinfo{year}{2005}).

\bibitem[{\citenamefont{Collin}(1991)}]{Collin}
\bibinfo{author}{\bibfnamefont{R.~E.} \bibnamefont{Collin}},
  \emph{\bibinfo{title}{Field Theory of Guided Waves}}
  (\bibinfo{publisher}{IEEE Press}, \bibinfo{year}{1991}),
  \bibinfo{edition}{2nd} ed.

\bibitem[{\citenamefont{Urzhumov et~al.}(2012)\citenamefont{Urzhumov, Chen,
  Bingham, Padilla, and Smith}}]{Urzhumov2012}
\bibinfo{author}{\bibfnamefont{Y.}~\bibnamefont{Urzhumov}},
  \bibinfo{author}{\bibfnamefont{W.}~\bibnamefont{Chen}},
  \bibinfo{author}{\bibfnamefont{C.}~\bibnamefont{Bingham}},
  \bibinfo{author}{\bibfnamefont{W.}~\bibnamefont{Padilla}}, \bibnamefont{and}
  \bibinfo{author}{\bibfnamefont{D.}~\bibnamefont{Smith}},
  \bibinfo{journal}{Phys.\ Rev.~B} \textbf{\bibinfo{volume}{85}},
  \bibinfo{pages}{054430} (\bibinfo{year}{2012}).

\bibitem[{\citenamefont{Rodr{\'\i}guez-Fortu–o
  et~al.}(2014)\citenamefont{Rodr{\'\i}guez-Fortu–o, Vakil, and
  Engheta}}]{Engheta2014}
\bibinfo{author}{\bibfnamefont{F.}~\bibnamefont{Rodr{\'\i}guez-Fortu–o}},
  \bibinfo{author}{\bibfnamefont{A.}~\bibnamefont{Vakil}}, \bibnamefont{and}
  \bibinfo{author}{\bibfnamefont{N.}~\bibnamefont{Engheta}},
  \bibinfo{journal}{Phys.\ Rev.\ Lett.} \textbf{\bibinfo{volume}{112}},
  \bibinfo{pages}{033902} (\bibinfo{year}{2014}).

\end{thebibliography}

\end{document}